\documentclass[prd,twocolumn,aps,amsmath,amssymb,nofootinbib,preprintnumbers]{revtex4}
\usepackage{graphicx,epsfig}
\usepackage{bm}
\usepackage{amsmath}
\usepackage{amsfonts}
\usepackage{verbatim}

\newcommand{\beq}{\begin{equation}}
\newcommand{\eeq}{\end{equation}}
\newcommand{\bea}{\begin{eqnarray}}
\newcommand{\eea}{\end{eqnarray}}

\begin{document}
\title{Tuned MSSM Higgses as an inflaton}

\author{Arindam Chatterjee$^{1}$}
\author{Anupam Mazumdar$^{2,3}$}

\affiliation{$^{1}$~Physikalisches Institut der Universit\"a{}t Bonn, Nu\ss{}allee 12, Bonn, 53115, Germany\\
$^{2}$~Physics Department, Lancaster University, LA1 4YB, UK\\
$^{3}$~Niels Bohr Institute, Copenhagen University, Blegdamsvej-17, Denmark}


\begin{abstract}
We consider the possibility that the vacuum energy density of the MSSM (Minimal Supersymmetric Standard Model)
flat direction condensate involving the Higgses $H_1$ and $H_2$ is responsible for 
inflation. We also discuss how the finely tuned Higgs potential at high vacuum expectation values can realize {\it cosmologically} flat 
direction along which it can generate the observed density perturbations, and after the end of inflation -- the coherent 
oscillations of the Higgses  reheat the universe with all the observed degrees of freedom, without causing any problem for 
the electroweak phase transition.
\end{abstract}

\maketitle

\section{Introduction}

Inflationary paradigm has been extremely successful from observational
point of view as it is responsible for stretching the initial
perturbations to the present Hubble scales~\cite{WMAP}.  Since all
relevant matter, such as the Standard Model (SM) degrees of freedom, has
to be created after the end of inflation, it is obvious that the
inflaton vacuum cannot be arbitrary, for a review on inflation see~\cite{Infl-rev}. Inflation may
occur at different scales and at different vacua, but the last phase
of inflation must create all matter and observed perturbations in the
cosmic microwave background (CMB) radiation~\cite{Infl-rev}.

In order to achieve this one has to realize inflation within a {\it visible} sector where all the couplings are well known. 
The MSSM (Minimal Supersymmetric Standard Model) inflation is one such example where
inflation is driven by {\it gauge invariant} combinations of $D$-flat directions~\cite{AEGM,AKM,AEGJM}. 
The inflaton decays directly into the SM quarks and leptons, thus creating the baryons~\cite{AFGM}, 
and also the cold dark matter particles--as one of the lightest supersymmetric
particle (LSP)~\cite{ADM,ADS}. These models of inflation also solve the generic moduli problem-- with a TeV mass
moduli field is typically heavy during inflation, and thus never get displaced from
their minima, or their coherent oscillations die-off during inflation~\cite{AEGM,AEGJM,AKM}~\footnote{Inflation 
may as well proceed in many {\it hidden} sectors beyond the SM, and 
at sufficiently high scales.  The main obstacle such models face is how to reproduce the SM baryons and the observed 
abundance of dark matter~\cite{Infl-rev} from completely arbitrary scalar fields with couplings usually set by hand. In 
order to construct such hidden sector models we must know all the inflaton couplings to {\it hidden} and {\it visible} 
matter, and a proper embedding of inflation with a UV complete theory. One such unique model
has been constructed within type IIB string theory, where it was found that all the inflaton 
energy is transferred to exciting the {\it hidden} matter~\cite{Cicoli}, and the universe could be 
prematurely dominated by hidden sector dark matter.}.

In this paper our aim will be to show that under some fine-tuned conditions the MSSM Higgses can also drive inflation, their slow rolling
can create seed perturbations for the CMB, and excite all the relevant matter for the success of cosmology in the visible sector. 
Although the Higgs potential would require fine tuning in maintaining the flatness at the scale of inflation determined by the VEV, 
$10^{14}$~GeV, but at the weak scale due to the running of the masses and couplings the Higgs can facilitate an electroweak 
phase transition within a large parameter space of the MSSM. 

In section II, we will discuss the Higgs potential, in section III, we will discuss how inflation can be driven near the 
point of inflection, and describe the parameter space which leads to a successful cosmology, in section IV, we will
discuss the issue of fine tuning and the electroweak symmetry breaking.

\section{The Potential}

Let us begin by considering the $\mu$ term of the MSSM superpotential.
Besides MSSM superpotential, it is also possible to have 
non-renormalizable terms of the following form in the superpotential~\cite{DRT,MSSM-rev}~\footnote{The $\mu$-term has 
been considered in past to generate the density perturbations, see~\cite{Kasuya-M}.},
\beq 
\mathcal{W} = \mu {\bf H_{1}}. {\bf H_{2}} +\frac{\lambda_k}{k} \frac{\left({\bf H_{1}}.
{\bf H_{2}}\right)^k} { M^{2k-3}_{P}},
\label{supot}
\eeq
where ${\bf H_{1}}$ and ${\bf H_{2}}$ are Higgs superfields, $M_P$ denotes the
reduced Planck mass, equals to $ 2.4 \times 10^{18}$ GeV. 
The scalar Higgs fields, $H_{1}, H_{2}$, which give masses to the 
down-type and up-type quarks respectively, constitutes the scalar 
components of ${\bf H_{1}}$ and ${\bf H_{2}}$ respectively. In the above equation
$k \geq 2$. For simplicity, we begin by considering one such non-renormalizable 
term at a time.

We then obtain the scalar potential along the $H_{1} H_{2}$ D-flat direction,
\bea\label{soft-pot}
\tilde{V} (\varphi,\theta) & = & \frac{1}{2} m^{2}(\theta)\varphi^{2}
   +(-1)^{(k-1)} 2 \lambda_{k}^{'} \mu \cos ((2k-2)\theta) \varphi^{2k} \nonumber \\
  & + & 2 \lambda_{k}^{'2} \varphi^{2(2k-1)},
\eea
where $\phi = |\phi| e^{i \theta}$, $\varphi  =   \sqrt{2} |\phi|$, and 
\bea
H_{1} & = & \frac{1}{\sqrt{2}}(\phi, 0)^{T}, \\
H_{2} & = & \frac{1}{\sqrt{2}}(0, \phi)^{T}, \\
m^{2}(\theta) & = & \frac{1}{2}(m_{1}^{2} + m_{2}^{2}  
               +  2 \mu^{2}-2 B \mu \cos 2\theta ), \\ 
\lambda^{'}_{k}& = & \frac{\lambda_k}{2^{(2k-1)}M^{2k-3}_{P}}\,.
\label{lambda-k} 
\eea

For simplicity here we assume $\mu$ and $ B$ to be real. This choice
is compatible with the experimental constraints, mainly from the EDM 
measurements and is well motivated, especially in the context  
of weak scale supersymmetry \cite{CPVREV}. 

The potential is bounded from below for $ m^{2} \geq \mu^{2}$, 
and the minimum is $0$ at $\varphi = 0 $. However, it is also possible to
have local extrema. For any $\lambda_{k}$, we have
\bea
\frac{\partial \tilde{V}}{\partial \theta} & = &
 2 B \mu \sin (2\theta) \varphi^{2} \nonumber \\
& - &  (-1)^{k-1} 2(2k-2)
\lambda_{k}^{'} \mu \sin((2k-2) \theta) \varphi^{2k}, \label{dertheta1}\\
\frac{\partial^{2} \tilde{V}}{\partial \theta^{2}} & = &
 4 B \mu \cos (2\theta) \varphi^{2} \nonumber \\
 & - &  (-1)^{k-1} 2(2k-2)^{2}
\lambda_{k}^{'} \mu \cos((2k-2) \theta) \varphi^{2k}. \nonumber \\
\label{dertheta2}
\eea
Considering principal values of $\theta$, $\tilde{V} (\varphi, \theta)$, which
may posses a secondary minimum for $\theta \in \{0, \pm\frac{\pi}{2}\}$.
To illustrate further, with $\theta = 0 $, we have
\bea
\frac{\partial V(\varphi)}{\partial \varphi} & = & m_{0}^{2} \varphi 
+ (-1)^{k-1}4k \lambda^{'}_{k} \mu \phi^{2k-1} \nonumber \\
&  &+ 4(2k-1) \lambda^{'2}_{k} \varphi^ {4k-3}, \label{derphi1}\\
\frac{\partial^{2} V(\varphi)}{\partial \varphi^{2}} & = & m_{0}^{2} 
+ (-1)^{k-1} 4k(2k-1) \lambda^{'}_{k} \mu  \phi^{2(k-1)}\nonumber \\
&  &  + 4(2k-1)(4k-3) \lambda^{'2}_{k} \varphi^ {4(k-1)}, 
\label{derphi2}
\eea
where, $V (\varphi) = \tilde{V}(\varphi, 0), ~m_{0} = m(\theta= 0)$.
The necessary conditions for the existence of the secondary local minimum,  
required by the reality of $\varphi$, are given by
\beq
k^{2} \mu^{2} \geq m_{0}^{2} (2k-1),
\label{ineq}
\eeq
and also the coefficient of $\varphi^{2k-1}$ in Eq.~(\ref{derphi1})
must be negative. At this minimum, we have 
\beq 
\varphi = \varphi_0
\sim \left(m_0 M^{2k-3}_{P}\right)^{1/2k-2} \ll M_{P} 
\eeq
The curvature of the potential along the radial
direction is $\, m^2_{0}$, and the potential reduces to: 
\beq
V \sim m_0^2\varphi_0^2 \sim
m_{0}^2\left(m_{0} M^{2k-3}_{P}\right)^{2/(2k-2)}\,.
\eeq
This is the situation when the flat direction can get trapped in the
false minimum $\varphi_0$. If its potential energy, $V$, dominates the
total energy density of the universe, a period of inflation is
obtained. However, one finds that $H_{\rm inf}\sim {m_0 \varphi_0 }/{ M_{P}} \sim
m_0 \left(m_0/{M_{P}}\right)^{1/(2k-2)} \ll m_0$. This implies that the potential is
too steep at the false minimum and $\phi$ cannot climb over the
barrier which separates the two minima just by the help of quantum
fluctuations during inflation. The situation is essentially the same
as in the old inflation scenario~\cite{GUTH} with no
graceful exit from inflation.

\section{Inflection point inflation}

This issue can be resolved if inflation occurs around an inflection point 
instead \cite{AKM,Lyth,AEGJM,EMS}. The inflection point can be obtained 
if the following condition is satisfied,  
\beq
m^{2}_{0}= \frac{k^{2} \mu^{2}}{(2k-1)} + \tilde{\lambda}^{2}, 
\label{cond1}
\eeq
where $\tilde{\lambda}$ is the tuning required to lift the potential~\footnote{When 
$ k^{2} \mu^{2} = (2k-1) m^{2}_{0}$, the potential accommodates a saddle point. In this paper
we will follow the analysis of inflection point. }. If this relation holds good at the appropriate 
energy scale, an inflection point exists at $ \varphi_0$, where, 
\bea\label{finetuning}
\lambda^2 & = & \frac{2k-1} {8(k-1)^{2}}\frac{\tilde{\lambda}^{2}}{\mu^2 k^2}
 =  \frac{\tilde{\lambda} ^{2} m_{0}^{-2}}{8(k-1)^{2}}, \\
\varphi_0 & = & \left(\frac{k |\mu||\lambda^{'}_k|^{-1}}{2(2k-1) }\right)^{1/(2k-2)} 
\left(1-\lambda^{2}\right)+ \mathcal{O}(\lambda^4), \nonumber\\
& = & \left(\frac{m_{0} |\lambda^{'}_k|^{-1}}{2\sqrt{2k-1}}\right)^{1/2k-2} (1-\lambda^{2})
+ \mathcal{O}(\lambda^4).
\eea

Around the inflection point $\varphi_0$, the potential for the inflaton may be 
expanded in a Taylor series,
\beq
V(\varphi)=V_{0} + \alpha_{1}(\varphi -\varphi_0)+\frac{1}{3!} \alpha_{3}
(\varphi-\varphi_0)^{3}+ ... ~.
\eeq
where, $V_{0} = V(\varphi_0)$, $ \alpha_{1}= \dfrac{\partial V(\varphi_)}{\partial \varphi}|_{\varphi = \varphi_0}$ 
and $ \alpha_{3}= \dfrac{\partial^{3} V(\varphi_)}{\partial \varphi^{3}}|_{\varphi = \varphi_0}$.
Here we have assumed that,
\bea
|\alpha_{1}| \gg \left|\dfrac{d^{n} V}{d\varphi^{n}}\right|_{\varphi=\varphi_0}
|\varphi_e - \varphi_0|^{n-1}, \\
|\alpha_{3}| \gg \left|\dfrac{d^{n} V}{d\varphi^{n}}\right|_{\varphi=\varphi_0}
|\varphi_e - \varphi_0|^{n-3},
\eea
where $ n \geq 4 $ and $\varphi_e$ denotes $\varphi$ at the end of slow-roll regime.
In terms of the relevant parameters, we also have,
\bea
V_0 & = &  \mu^{2} k \frac{(k-1)^2}{(2k-1)^2}\left(\frac{k |\mu||\lambda^{'}_{k}|^{-1}}{2(2k-1)}\right)^{1/k-1}+\mathcal{O}(\lambda^2), \nonumber \\
& = & \frac{(k-1)^{2}m_{0}^{2}}{k(2k-1)}\varphi_{0}^{2}+\mathcal{O}(\lambda^2), \\
\alpha_{1} & = & \tilde{\lambda} ^{2} \left(\frac{k|\mu||\lambda^{'}_{k}|^{-1}}{2(2k-1)}\right)^{1/2k-2}+\mathcal{O}(\lambda^4)\nonumber \\
& = & 8(k-1)^{2}\lambda ^{2}m_{0}^{2}\varphi_{0} + \mathcal{O}(\lambda^4), \\
\alpha_{3} & = & 8(k-1)^{2} ~\frac{m_{0}^{2}}{\varphi_{0}}+ \mathcal{O}(\lambda^2).
\eea 

From now on we only keep the leading order terms in all the expressions. Note that the success of 
electroweak symmetry breaking requires the $\mu$-term and the soft breaking 
mass to be the same order of magnitude, $m_{0}\sim \mu \sim {\cal O}(100)$~GeV.

The slow-roll parameters, in the vicinity of the inflection point, are given by,
\bea
\epsilon(\varphi) & = & \frac{1}{2}\left(\frac{V'(\varphi)}{V(\varphi)}\right)^{2}
= \frac{M_{P}^{2}}{2 V_{0}^{2}} (\alpha_{1} + \frac{\alpha_{3}}{2} (\varphi -\varphi_0)^{2})^2 , \\
\eta(\varphi) & = & {M_{P}^{2}}\frac{V^{''}(\varphi)}{V(\varphi)} = 
{M_{P}^{2}}\frac{\alpha_3}{V_0}(\varphi-\varphi_0), \\
\xi(\varphi) & = &  M_{P}^{4} \frac{V'(\varphi) V^{'''}(\varphi)}{V(\varphi)^{2}} \nonumber\\
& = & M_{P}^{4} \frac{\alpha_{3} }{V_{0}^{2}}(\alpha_{1} + \frac{\alpha_{3}}{2} 
(\varphi -\varphi_0)^{2}).
\eea
The Hubble expansion rate during the slow-roll inflation is,
\beq
H_{inf}\simeq \sqrt{\frac{V_{0}}{3 M_{P}^{2}}} = \frac{k-1}{\sqrt{3 k(2k-1)}} 
\frac{m_{0} \varphi_{0}}{M_P}.
\eeq
For $\varphi_{0}\sim 10^{14}$~GeV, a typical VEV at the inflection point in our context~\footnote{The initial conditions for low scale
inflation have been discussed in Refs.~\cite{AFM,Initial, Initial2}.} 
\begin{equation}\label{Hinf}
 H_{inf} \sim \frac{m_{0}\phi_{0}}{M_{P}}\sim 10^{-1}~{\rm GeV}, 
\end{equation}
Inflation ends when $|\eta| \simeq 1$. The interval suitable for
inflation is given by, 
\beq \label{end}
\frac{|\varphi_0-\varphi|}{\varphi_0} \sim
\left(\frac{\varphi_0} {8k(2k-1)M_{P}}\right)^{2}\,.
\eeq


\begin{figure}[t]
\epsfig{file=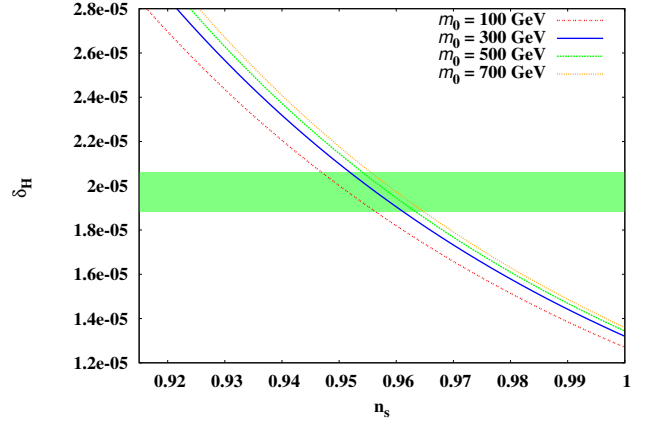,width =\linewidth, height = 6 cm}
\caption{$\delta_{H}$ and $n_s$ have been plotted for different values of $m_0$ and at the inflection point VEV, $\varphi_{0}\sim 3\times 10^{14}$~GeV.
 We have used $k=2$ in the superpotential, we have taken  $\lambda^{'}_{2} M_{P} = 1.4 \times 10^{-8}$, see Eq.~(\ref{lambda-k}).
 The green band denotes  $2\sigma$ allowed band of $\delta_H$~\cite{WMAP}. Although the splitting between these curves are not so sensitive to the inflaton mass, 
varying $\lambda^{'}_{2}$ it is possible to span the complete range in the $n_s$-$\delta_H$ plane.}
\label{dhns1}
\end{figure}


\begin{figure}[t]
\epsfig{file=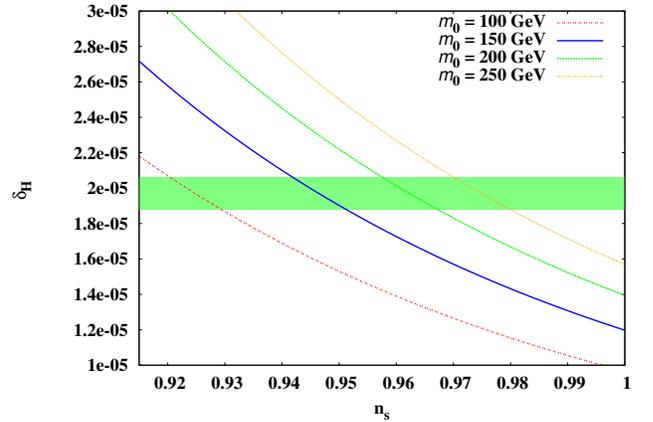, width =\linewidth, height = 6 cm }
\caption{$\delta_{H}$ and $n_s$ have been plotted for different values of $m_0$ and at the inflection point VEV, $\varphi_{0}\sim 3\times 10^{14}$~GeV. 
We have used $ k=3$ in the superpotential, we have taken $\lambda^{'}_{3} M_{P}^{3} = -0.71$, see Eq.~(\ref{lambda-k}). 
The green band denotes $2\sigma$ allowed band of $\delta_H$~\cite{WMAP}.}
\label{dhns2}
\end{figure}


\begin{figure}[t]
\epsfig{file=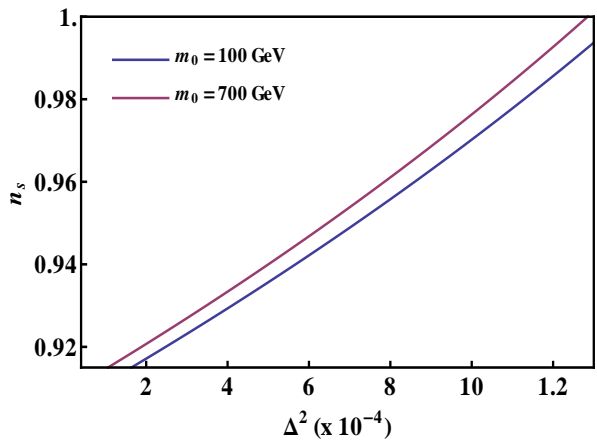,width =\linewidth, height = 6 cm}
\caption{$n_s$ has been plotted against vs $\Delta^{2}$ for different 
values of $m_0$ for  $k=2 $ case with $\lambda^{\prime}_{2}M_{P}= 1.4 \times 10^{-8}$.}
\label{nsdelta1}
\end{figure}


The amplitude of density perturbation generated is given by \cite{Lyth,AEGJM,EMS},
\begin{eqnarray}
\delta_H \simeq \frac{1}{5 \pi} \sqrt{\frac{2}{3} 2k(2k-1)}
(2k-2) \left( \frac{m_0 M_{P}}{\varphi_{0}^{2}} \right) \,\nonumber \\
\frac{1}
{\Delta^2} \sin^{2}[\mathcal{N}_{\rm COBE} \sqrt{\Delta^2} ], 
\end{eqnarray}
and the spectral index for the scalar perturbation is given by \cite{Lyth,AEGJM,EMS}, 
\beq
n_s = 1- 4 \sqrt{\Delta^{2}} \cot[\mathcal{N}_{\rm COBE} 
\sqrt{\Delta^{2}}],
\eeq
where,
\beq
\Delta^2 = 32 k^{2} (2k-1)^{2} \lambda^{2}{\cal N}_{\text{COBE}}^{2} 
\left(\frac{M_P}{\varphi_0}\right)^{4}. 
\eeq

${\cal N}_{\text {COBE}}$ is the number of e-foldings between the time when the 
observationally relevant perturbations were generated till the end of inflation
and is given by  ${\cal N}_{\text {COBE}} \simeq 66.9+(1/4)\ln(V(\varphi_{0})/M_{P}^{4})\sim 45$~\cite{Leach},
for either $k=2, 3$, provided that the universe thermalizes within one-Hubble time after the end 
of inflation--see the discussion below.

So far the expressions hold good for any general $k$. Now let us consider two simple examples.

\begin{itemize}
\item{$k=2$ case:\\
Let us consider the lowest order non-renormalizble term, with $k=2$.
Using the general formulation we described, we plot $\delta_H$ vs $n_s$ 
in Fig.~[\ref{dhns1}]. In the context of low energy supersymmetry, with 
inflaton masses of about $100-1000$~GeV, it is possible to obtain the required
density perturbations and the spectral index, consistent with the WMAP data 
\cite{WMAP}, as indicated by the figure, for the VEV around $\varphi_{0}\sim 10^{14}$~GeV.

However, the coefficient $\lambda_2$, which naturally may be expected to be of order ${\cal O}(1)$, 
needs to be of order $10^{-8}$, or to be more specific $\lambda_{2}^{\prime}M_{P}\sim 1.4\times 10^{-8}$, see Eq.~(\ref{lambda-k}), 
instead. We note that there are other dimension 5 operators, even allowed by R-parity, which needs 
similar suppression to satisfy constraints from the proton deacy. 
These are $QQQL$ and $\bar{U} \bar{U} \bar{D} \bar{E}$, see, for instance, \cite{proton}.}

\item{$k=3$ case:\\
On the other hand, assuming $\lambda_{2} \approx 0$ and 
considering only $k=3$ term may lead to a satisfactory result. It is possible to match the CMB observations 
for $\delta_H$ and $n_s$ for $\lambda_{3}\sim {\cal O}(1)$, or to be more specific $\lambda_{3}^{\prime}M_{P}^{3}\sim -0.71$, see Eq.~(\ref{lambda-k}). 
We demonstrate this in Fig.~[\ref{dhns2}], where we assumed $\lambda_{2}=0$~\footnote{Strictly speaking we can keep both $\lambda_{3}\sim {\cal O}(1)$
and  $\lambda_{2}\sim 10^{-8}$ at $\varphi_0 \sim 10^{14}$ GeV.}. }
\end{itemize}

Since inflation occurs at low scale, the tensor to scalar ratio remains small.
From Fig.~ [\ref{nsdelta1}], we see that in order to obtain the spectral index, consistent with
the recent data, we need 
\begin{equation}\label{finetune-1}
\Delta^{2} \sim 10^{-3}~~\Longrightarrow~~ \lambda \sim 
10^{-11}\,. 
\end{equation}
Although the figure assumes $k=2$, a similar plot may also
be obtained with $k=3$. Further note that $\lambda$, see Eq.~(\ref{finetuning}), remains of the same order for both 
cases, $k =2$ and $k =3$~\footnote{Although we have illustrated with $\theta =0 $, a similar calculation
may be performed with $\theta = \pm \pi/2$, with appropriate choice of
parameters.}.

After inflation the coherent oscillations of the Higgses excite
the MS(SM) degrees of freedom and reheat the universe.The thermalization time scale is much shorter than one Hubble time.
The Higgses start coherent oscillations around their minimum denoted by $\varphi=0$, which is the point of 
 {\it enhanced gauge symmetry}--where the entire SM gauge symmetry is restored. 
When ever the inflaton passes through the point of enhanced gauge symmetry, 
the zero mode of the inflaton condensate excites the massless
modes of MSSM via non-perturbative phenomena~\cite{TB,KLS}, and~\cite{AEGJM,AFGM}, known as  {\it instant preheating}~\cite{instant,KLS}. 
These are the massless modes which couple to the inflaton directly, for instance the degrees of freedom corresponding  to 
$SU(3)_{C}\times SU(2)_W\times U(1)_Y$. At VEVs away from the minimum, the same modes become heavy and therefore it becomes
kinematically unfavorable to excite them. 

The actual process of excitation depends on how strongly the adiabaticity condition for the time 
dependent vacuum is violated for the inflaton zero mode~\cite{instant,KLS}. Nevertheless, draining the inflaton energy is quite efficient, as shown in a particular MSSM flat direction inflaton~$LLe$, where
nearly $10\%$ of the inflaton energy density gets transferred to the relativistic species. It takes near about 120 oscillations 
to reach the full {\it chemical} and {\it kinetic} equilibrium via processes requiring $2\leftrightarrow 2$ and $2\leftrightarrow 3$
interactions~\cite{AFGM}, given that one Hubble time corresponds to roughly 1000 oscillations. Note that the dynamical 
forms of the potentials for $LLe$ and $H_{1}H_{2}$ are very similar.
Since $H_{1}H_{2}$ couples all the MS(SM) degrees of freedom, it is expected that the inflaton energy density would decay at a similar rate and thermalization 
time scale would be very similar to the case of $LLe$, which couples mainly to $SU(2)_{W}\times U(1)_{Y}$ degrees of freedom. In any case,
due to the hierarchy in the scales, $m_{0}\sim 1 $~TeV, and $H_{inf}\sim 10^{-1}$~GeV, the thermalization time scale will be well short of
$1000$ inflaton oscillations, which marks the one Hubble time after inflation. For all practical purposes reheating and thermalization will be
over instantly.  The final reheat temperature will be given by:
\begin{equation} \label{Trh}
T_{\rm rh} \simeq \left(\frac{30}{\pi^2 g_*}\right)^{1/4} \rho_0^{1/4} \simeq 2 \times 10^8 ~ {\rm GeV} \, ,
\end{equation}
where we have used $g_* = 228.75$ (all degrees of freedom in MSSM) and $\rho_{0} = (4/15)m_{0}^{2}\phi_{0}^{2}$~\cite{AFGM}.


\section{Fine tuning and Electroweak symmetry breaking}

Note that a fine tuning ($\lambda$) of order $10^{-11}$ is 
required for inflation to occur, see Eqs.~(\ref{finetuning},~\ref{finetune-1}). However, 
in our case the tuning involves only the MSSM parameters, although at a high scale. 
At one-loop level, these parameters evolve by the respective renormalization 
group equations (RGEs), as in the  MSSM, see~\cite{Drees}. 

Ignoring the Yukawa couplings for the first two generations of fermions, the
RGEs for $ m_{1}, ~m_{2}, ~ B, ~\mu$ are given by~\cite{Drees},
\bea
\frac{d \mu}{d \mathcal{E}} & = & \frac{\mu}{16 \pi^2} 
\left( 3 f_{t}^{2} + 3 f_{b}^{2} + f_{\tau}^{2} -3 g_{2}^{2}- g_{Y}^{2} \right), \\
\frac{d B}{d \mathcal{E}} & = & \frac{1}{8 \pi^2} 
\left(- 3 f_{t}^{2}A^{t} - 3 f_{b}^{2}A^{b} - f_{\tau}^{2}A^{\tau} + 3 g_{2}^{2} M_{2} \right. \nonumber \\ 
& &  \left. + g_{Y}^{2} M_{1} \right),\\
\frac{d m_{1}^{2}}{d \mathcal{E}} & = & \frac{1}{8 \pi^2} 
\left( 3 f_{b}^{2}  (m_{1}^{2} + m_{\tilde{Q}_3}^{2} + m_{\tilde{b}_R}^{2} + |A^{b}|^{2})
 \right. \nonumber \\ & & \left.+ f_{\tau}^{2} (m_{1}^{2} + m_{\tilde{l}_3}^{2} 
+ m_{\tilde{\tau}_R}^{2} + |A^{\tau}|^{2}) -3 g_{2}^{2}|M_{2}|^{2}
\right. \nonumber \\ & & \left.- g_{Y}^{2}|M_{1}|^{2} 
 -\frac{1}{2} g_{Y}^{2} S_{Y} \right),\\
\frac{d m_{2}^{2}}{d \mathcal{E}} & = & \frac{1}{8 \pi^2} 
\left( 3 f_{t}^{2} (m_{2}^{2} + m_{\tilde{Q}_3}^{2} + m_{\tilde{t}_R}^{2} + |A^{t}|^{2})
\right. \nonumber \\ & & \left.  -3 g_{2}^{2}|M_{2}|^{2}- g_{Y}^{2}|M_{1}|^{2} 
+\frac{1}{2} g_{Y}^{2} S_{Y} \right),
\eea
where $f_i$s denote the Yukawa couplings for the i-th 
fermion species, $g_{2}$ and $g_Y$ denote the $SU(2)_W$ and $U(1)_Y$ gauge 
couplings respectively, $m_i$s and $A^i$s denote the soft supersymmetry breaking mass 
terms and the tri-linear terms respectively. $M_i$s denote the soft supersymmetry breaking 
gaugino masses, and, 
\beq
S_Y = \frac{1}{2}\Sigma_i Y_{i} m_{i}^{2},
\eeq 
where, the sum runs over all the scalar fields in MSSM, with masses $m_i$ and 
hypercharges $Y_i$. Also, all the other MSSM parameters involved in the above 
equations evolve simultaneously with $\mathcal{E}$ according to the respective RGEs. 

Note that in our analysis, see Eq.~(\ref{soft-pot}), we have not considered any 
soft supersymmetry breaking term of the form $ A \lambda^{'}_{k}\varphi^{2k}$ 
in the Higgs sector. While this reduces the number of parameters involved, phenomenologically these terms 
are not forbidden. However, the origin of such a term may be determined by  
specific theories of supersymmetry breaking. Even if such a term is absent in the 
Higgs sector, similar terms involving Higgs fields and the sfermions can 
give rise to these terms at next to the leading order. 

Considering a term of the form $ A \lambda^{'}_{k}\varphi^{2k}$ will lead to 
a straight forward modification of Eq.~(\ref{cond1})
removing any tuning between $m_0$ and $\mu$. To illustrate, with $\theta = 0$ 
and $\lambda_{k}^{'} A < 0$, Eq.~(\ref{cond1}) will be modified to,
\beq
m^{2}_{0}= \frac{k^{2} ((-1)^{k-1} 2\mu+A)^{2}}{4(2k-1)} + \tilde{\lambda}^{2}.
\eeq 

\begin{figure}[t]
\epsfig{file=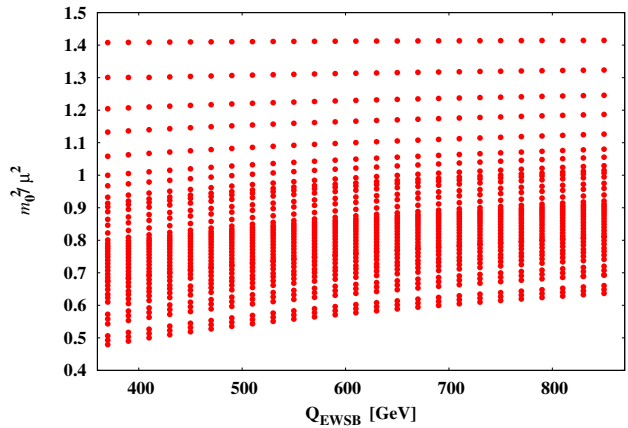, width =\linewidth, height = 6 cm}
\caption{A sample plot where the ratio $m_{0}^2/\mu^2$, see Eq.~(\ref{cond1}), for $k =2$, has been evaluated at 
the EWSB scale. The corresponding value at a high scale, $\varphi_{0}\sim 10^{14}$ GeV, is set to $4/3$, see Eq.~(\ref{cond1}),
with an accuracy of 0.1\%. The RGE accuracy in $SuSpect$~\cite{Kneur} is about 0.01\%.}
\label{ftuning1}
\end{figure}

We demonstrate in Fig.~(\ref{ftuning1}), that various points in the parameter 
space, which give the tuning required at the scale of the VEV $\varphi_0$, 
do not satisfy any particular relation when evolved to the EWSB scale. 
The figure also shows that successful EWSB is possible over a large 
parameter space in spite of the tuning. We use the publicly available code $SuSpect$
to evolve the RGEs~\cite{Kneur}, modifying it a little to cater our needs.


\begin{figure}[t]
\epsfig{file=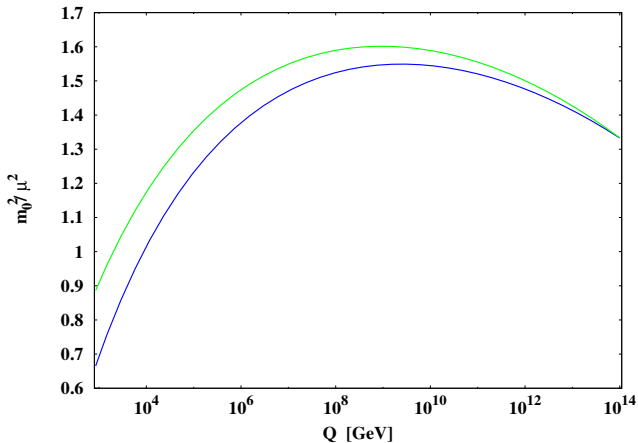, width =\linewidth, height = 6 cm}
\caption{The ratio $m_{0}^2/\mu^2$ for $k =2$, has been evolved from
$10^{14}$ GeV to the EWSB scale (chosen to be 850 GeV). The green line
and the blue line correspond to $m_0 = 323.4 $ GeV and $m_0 = 354 $ GeV 
at $10^{14}$ GeV respectively. The ratio at the high scale 
($10^{14}$ GeV) is set to 4/3, see Eq.~(\ref{cond1}), with an accuracy of 
0.1\%. The RGE accuracy in $SuSpect$~\cite{Kneur} is about 0.01\%.}
\label{ftuning2}
\end{figure}

In Fig.~ (\ref{ftuning2}) we evolve the ratio $m_{0}^2/\mu^2$
from a high scale, typical for $\varphi_0$, to the EWSB scale. 
The plot demonstrates that the ratio evolves with scale, and 
differs by about a part in 10 at the EWSB scale. 


Thus the fine tuning between the parameters, as given by Eq.~(\ref{cond1}) may be satisfied dynamically. The 
relevant parameters, while assigning arbitrary ($\sim 1$) values of $\lambda$ at a low scale, 
can still satisfy Eq.~(\ref{cond1}) at the scale of inflection point $\varphi_0\sim 10^{14}$~GeV.



\section{Conclusion}

To summarize, we have shown that a combination of MSSM Higgs 
fields can be the inflaton. A sufficient number of e-foldings and the 
right amount of density perturbations are obtained if, apart from the $\mu$ term, 
the ${H_1} {H_2}$ (D-)flat direction is lifted by 
$\left( H_{1}.H_{2}\right)^k$ operator, for $k =2$, and $k=3$. 
However, for $k=2$, the coefficient $\lambda_{2}$ needs to be suppressed. 
The model predicts low scale inflation,
non-detectable gravity waves, and a slight departure from scale
invariance. Further note that supergravity
effects are negligible for $H_{\rm inf}\ll m_{0}$ and therefore do not
spoil the predictions~\cite{AEGM,AEGJM}. More importantly, there will be 
no moduli problem in our model. Since $H_{\rm inf} \ll m_0$, all moduli 
will settle at their true minimum during inflation.

The salient feature of the present model is that the inflaton is now 
identified with the MSSM Higgs fields. Therefore its
properties could be determined independently of cosmology by particle
physics experiments. In particular the condition in Eq.~(\ref{cond1}) for 
having a flat potential amounts to a relationship among the parameters of 
the Higgs sector which can be determined at the LHC. Further work is 
required to establish these connections.

{\it Acknowledgments-} We wish to thank Rouzbeh Allahverdi, Manuel Drees and Jean-Loic Kneur 
for helpful discussions.


\end{document}